\documentclass[conference]{IEEEtran}
\IEEEoverridecommandlockouts
\usepackage{cite}
\usepackage{amsmath,amssymb,amsfonts}
\usepackage{graphicx}
\usepackage{textcomp}
\usepackage{xcolor}
\usepackage{booktabs}
\usepackage{bigstrut}
\usepackage{algorithm}
\usepackage[noend]{algpseudocode}
\usepackage{xspace}
\usepackage{comment}
\usepackage{multirow} 
\usepackage{threeparttable}
\usepackage{url}
\algrenewcommand\algorithmicindent{0.7em}
\usepackage{amsmath}
\def\BibTeX{{\rm B\kern-.05em{\sc i\kern-.025em b}\kern-.08em
    T\kern-.1667em\lower.7ex\hbox{E}\kern-.125emX}}
\begin{document}

\title{GAMA: High-Performance \underline{G}EMM \underline{A}cceleration on AMD Versal \underline{M}L-Optimized \underline{A}I Engines \vspace{-3mm}}

\author{\IEEEauthorblockN{Kaustubh Manohar Mhatre}
\textit{Arizona State University}\\
Tempe, USA \\
kmhatre@asu.edu
\and
% \linebreakand % Break to the next row
\IEEEauthorblockN{Endri Taka}
\textit{The University of Texas at Austin}\\
Austin, USA \\
endri.taka@utexas.edu
\and
\IEEEauthorblockN{Aman Arora}
\textit{Arizona State University}\\
Tempe, USA \\
aman.kbm@asu.edu}

% High frequency words 
\newcommand{\Pname}{GAMA}
%changed to aie. added a footnote on page 1
\newcommand{\aieml}[1][]{AIE#1\xspace}
\newcommand{\bol}{\textit{BufferOptLevel }}
\newcommand{\gama}{GAMA }
\newcommand{\maxeva}{MaxEVA }
\newcommand{\red}{\textcolor{red}}
\newcommand{\rev}{\textcolor{black}}
\maketitle

% \setlength{\dbltextfloatsep}{0pt}
% \setlength{\floatsep}{0pt}
% \textfloatsep = 1\baselineskip plus 0.2\baselineskip minus 0.4\baselineskip

\begin{abstract}

General matrix-matrix multiplication (GEMM) is a fundamental operation in machine learning (ML) applications. 
We present the first comprehensive performance acceleration of GEMM workloads on AMD’s second-generation AIE-ML architecture, which is specifically optimized for ML applications. 
Compared to AI-Engine (AIE), AIE-ML offers increased compute throughput and larger on-chip memory capacity. 
We propose a novel design that maximizes AIE-ML memory utilization, incorporates custom buffer placement within the AIE-ML and staggered kernel placement across the AIE-ML array, significantly reducing performance bottlenecks such as memory stalls and routing congestion, resulting in improved performance and efficiency compared to the default AMD's compiler.
% compiler provided by AMD.
% We propose a novel design that maximizes internal memory utilization and incorporates custom buffer placement within the \aieml array, significantly reducing performance bottlenecks such as memory stalls, resulting in improved performance and efficiency compared to the default compiler provided by AMD. 
% \textcolor{red}{Add a line about staggered kernels placement}
We evaluate the performance benefits of our design at three levels: single AIE-ML, pack of AIE-ML's and the complete AIE-ML array. 
\gama achieves state-of-the-art performance, delivering up to 165 TOPS (85\% of peak) for int8 precision and 83 TBFLOPS (86\% of peak) for bfloat16 precision GEMM workloads. 
Our solution achieves 8.7\%, 9\%, 39\% and 53.6\% higher peak throughput efficiency
compared to the state-of-the-art AIE frameworks AMA, MAXEVA, ARIES and CHARM, respectively.

\end{abstract}


\section{Introduction}

AMD Versal is a heterogeneous architecture that integrates
Adaptive Intelligent Engines (AIE), programmable logic (PL)
and processing system (PS) which consists of CPUs. 
Such architectures can enable high performance for computationally
intensive applications such as machine learning (ML), which
are mainly dominated by General Matrix Multiplications
(GEMM). 
While prior research has extensively explored the
performance of GEMM on the first-generation AI Engine
(AIE) \cite{amd:2021:web:aie1_arch}, no existing work has investigated its performance
on the second-generation AIE-ML Engine \cite{amd:2023:web:aie2_arch}.

Compared to AIE, the AIE-ML\footnote{In this paper, for ease of reading, we use the word AIE to refer to AIE-ML.} architecture has larger AIE memory, 2x throughput for int8 precision, and supports
bfloat16 precision.
% [higher cascade bits and mem tiles?]
Design frameworks from prior work such as \maxeva \cite{taka:2023:fpt:maxeva}, AMA \cite{deng:2024:fpl:ama}, CHARM \cite{zhuang:2023:fpga:charm} exploit the architecture of AIE achieving significant peak performance. 
These frameworks, however, have architectural dependencies that limit their performance gains to AIE and make them not directly adaptable to AIE-ML. 

To efficiently exploit the available enhanced compute throughput of the AIE-ML, it is crucial to first maximize the single AIE performance
%data memory utilization 
and then ensure scaling to maximize the AIE array utilization. 
At the single AIE level, maximizing performance requires running largest possible GEMM sizes, which in turn requires maximizing AIE memory utilization to fit those sizes.
Naively maximizing AIE memory utilization often results in buffer allocation to the nearby AI engines, making it challenging to scale the design to the whole array.
Restricting the buffers to a single AIE to facilitate scaling often results in stalls due to the compiler’s default buffer placement strategies. 
No prior work focuses on buffer placement since none of them tries to maximize AIE memory utilization.

Another critical factor to maximize performance is to maximize the utilization of the AIE array. 
Due to limited PL-AIE interface resources and numerous ways of connecting the AIEs, scaling to the AIE array is non-trivial. 
For example, scaling a single AIE design to the complete array might not be feasible due to limited PLIO resources.
Generally, the single AIE design is scaled up to a group of AIEs which we call a ``pack", and then that pack is replicated across the AIE array.
Extensive design space exploration is required to find the optimal connectivity within a pack. 
Coupled with the large design space, automatic AIE kernel placement and buffer allocation at the full array level can result in compilation failures or lead to inefficient solutions with lower AIE array utilization. 
In particular, when scaling designs to the complete array, we observe compilation failures due to placement errors and routing congestion.
In this paper, we propose \textbf{GAMA}, a framework to address these challenges by systemically analyzing them and identifying solutions at three levels.
%dividing them into 3 levels and 
%the bottlenecks at each level. 

At the \textit{single AIE level}, we propose a custom buffer placement algorithm that maximizes the AIE memory utilization (100\% in some cases) and minimizes memory stalls while keeping the buffers in a single AIE. % to ensure efficient scaling.
%We tackle the problem of scaling by initially experimenting with a limited number of AIEs, ensuring that this compact configuration (pack) can be easily replicated to the complete entire array.
At the \textit{pack level}, we determine the optimal connectivity and pack size that guarantees scalability and performance, while taking into account the constraints posed by the limited PL-AIE interface resources.
%By applying optimizations from the single AIE level, such as maximizing AIE memory utilization and custom buffer placement to keep the buffers within the pack, we make our pack easily scalable.
Using custom buffer placement to keep buffers within the pack, we make our pack easily scalable.
Finally, at the \textit{array level} we replicate the pack by proposing a staggered placement pattern overcoming the compiler's routing congestion issues resulting in 94\% AIE array utilization. 
Furthermore, our custom buffer and kernel placement techniques,  alleviate the efforts required by the compiler, resulting in a significant reduction in compilation time.
%We put all these techniques into our framework called \gama. 

% The main contributions of this work are:
This work makes the following main contributions:
\begin{itemize}
    \item \gama is the first comprehensive study and implementation of GEMM acceleration on the AMD Versal AIE-ML architecture. \rev{It adopts a scalability-first approach that emphasizes scalability over the performance of individual kernels, promoting regularity in kernel mapping to facilitate straightforward replication.}
    % , leveraging \aieml's hardware characteristics such as PLIO interface, broadcasting for data reuse, cascade interface for partial sum transfers. By exploiting these, we minimize bottlenecks and maximize the efficiency of data movement across the AIE2 array. \
  
    \item \gama maximizes the internal memory usage of the AIE, achieving up to \textbf{100\%} utilization — the highest reported compared to all previous work. 
    It effectively addresses memory stall issues through a novel custom buffer placement algorithm, reducing stalls by an average of \textbf{12\%}, outperforming the standard compiler optimization strategies provided by AMD's compiler.
    % maximize the internal memory usage of AIE achieving up to 100\% utilization, which is the highest compared to all prior work. 
    % highest utilization levels across all prior work exceeding 97\%. 
    % It addresses memory stall issues using a novel custom buffer placement algorithm. This approach surpasses the standard compiler optimization strategies offered by Vitis reducing the stalls by an average 12\%. 

    \item 
    \gama achieves very high AIE array utilization of \textbf{94\%} while minimizing efficiency loss through a custom AIE kernel placement strategy that avoids PLIO routing congestion. 
    Furthermore, custom buffer and kernel placement methods significantly accelerate compilation time by \textbf{6x}.
    % This work achieves high \aieml array utilization (up to 94\%) and minimal efficiency loss due to our custom \aieml  placement that avoids PLIO routing congestion. Out custom buffer placement and \aieml  placement significantly accelerates the compilation time by x\%. \red{Need to add quantification}

    % \item Our proposed design supports GEMM workloads across multiple precisions, including int8 (input=int8; output=int8, int16, int32) and bfloat16 (input=bfloat16; output=bfloat16). 
    % We achieve up to \textbf{53.6\%} higher peak throughput utilization compared to prior work. 
    % We achieve up to 85\% utilization of peak throughput compared to prior work.

    \item This is the first work on GEMMs on Versal that supports multiple output precision (int8, int16, int32) for int8 inputs to accommodate various accuracy requirements and the first to show performance on bfloat16 on the AIE, achieving \textbf{86\%} peak throughput utilization of the chip. 
    For int8, we achieve up to \textbf{85\%} of the chip's peak throughput utilization, surpassing prior work's performance by up to \textbf{39\%}.   

    \item We open-source \gama for other researchers to leverage the framework at: \url{https://github.com/advent-lab/GAMA}.

   % \item Analytical model ?

  % \item We provide an analytical performance that can estimate performance for our designs. 
\end{itemize}

\section{Related Work}
% Table generated by Excel2LaTeX from sheet 'Results section '
\begin{table*}[t]
  \centering
  \caption{Comparing prior work focusing on GEMM acceleration on Versal architecture with the proposed GAMA framework}
    \begin{tabular}{c|c|p{5em}|c|p{5em}|c|p{5em}|c|c}
    \toprule
    \multicolumn{1}{c|}{ \textbf{Work}} & \multicolumn{1}{p{4em}|}{\textbf{AIE Version}} & \textbf{Precision (ip-op)} & \multicolumn{1}{p{5em}|}{\textbf{Pack Size}} & \multicolumn{1}{p{6em}|}{\textbf{Max AIEs used}} & \multicolumn{1}{p{6em}|}{\textbf{Cascade for reduction}} & \multicolumn{1}{p{5em}|}{\textbf{AIE kernel placement}} & \multicolumn{1}{p{5em}|}{\textbf{AIE buffer placement}} & \multicolumn{1}{p{6em}}{\textbf{AIE memory utilization}} \\
    \midrule
    \textbf{CHARM \cite{zhuang:2023:fpga:charm}} & AIE & fp32-fp32 & \multicolumn{1}{p{5em}|}{4\nolinebreak\xspace(fp32)} & 384/400\nolinebreak\xspace(96\%) & Yes   & \multicolumn{1}{p{5em}|}{Manual (Horizontal)} & Custom & {Low (75\%)} \\
    \midrule
    \rev{\textbf{AutoMM \cite{zhuang:2023:fpga:automm}}} & AIE & int16-int16,\newline{}int8-int8 & \multicolumn{1}{p{5em}|}{2\nolinebreak\xspace(int8)} & 288/400\nolinebreak\xspace(72\%),\newline{}192/400\nolinebreak\xspace(48\%) & Yes   & \multicolumn{1}{p{5em}|}{Manual (Horizontal)} & Custom & {Low (75\%)} \\
    \midrule
    \textbf{\maxeva\cite{taka:2023:fpt:maxeva}} & AIE  & fp32-fp32,\newline{}int8-int32 & 3,4   & 400/400\nolinebreak\xspace(100\%),\newline{}400/400\nolinebreak\xspace(100\%)& No    & \multicolumn{1}{p{5em}|}{Manual (Tetris)} & \multicolumn{1}{p{7em}|}{Auto (Compiler default)} & Low (75\%) \\
    \midrule
    \textbf{AMA \cite{deng:2024:fpl:ama}} & AIE  & fp32-fp16,\newline{}int8-int16 & 3,4   & 342/400\nolinebreak\xspace(85\%),\newline{}342/342\nolinebreak\xspace(85\%) & No  & \multicolumn{1}{p{5em}|}{Manual (Horizontal)} & \multicolumn{1}{p{7em}|}{Auto (Compiler default)} & Low (75\%) \\
    \midrule
    \textbf{RSN-XNN \cite{wang:2025:arxiv:rsn_xnn}}& AIE  & \multicolumn{1}{l|}{fp32-fp32} & \multicolumn{1}{c|}{4} & \multicolumn{1}{l|}{384/400\nolinebreak\xspace(96\%)} & Yes   & \multicolumn{1}{p{5em}|}{Manual (Horizontal)} & -     & Low (75\%) \\
    \midrule
    \textbf{ARIES \cite{zhuang:2025:fpga:aries}}& AIE  & fp32-fp32,\newline{}int16-int16,\newline{}int8-int8 & -       & 352/400\nolinebreak\xspace(88\%),\newline{}352/400\nolinebreak\xspace(88\%),\newline{}320/400\nolinebreak\xspace(80\%) & Yes   & \multicolumn{1}{p{5em}|}{Manual (Horizontal)} & Custom & Low (75\%) \\
    \midrule
    \textbf{\gama (Ours) }& AIE-ML  & int8-int32,\newline{}int8-int16,\newline{}int8-int8,\newline{}bf16-bf16 & 2-38  & 288/304\nolinebreak\xspace(94\%),\newline{}288/304\nolinebreak\xspace(94\%),\newline{}288/304\nolinebreak\xspace(94\%),\newline{}288/304\nolinebreak\xspace(94\%), & Yes   & Manual\newline{}(Staggered horizontal) & Custom & High (97\%)  \\
    \bottomrule
    \end{tabular}%
  \label{tab:prior_work_comparision}%
  \vspace{-6mm}
\end{table*}%

Since the introduction of AI engines by AMD \cite{amd:web:versal}, a wide variety of workloads have been deployed on them \cite{dong:2024:tcad:eq_vit, zhang:2022:fpl:hgcn, perryman:2023:ac:evaluation, singh:2023:ics:sparta, chen:2023:fpl:exploiting,yang:2023:iccad:aim,zhuang:2024:fpga:ssr,yemme:2023:ijcnn:scalable}.
Additionally, several frameworks have mapped and analyzed the performance of GEMM on these AI engines.
Table \ref{tab:prior_work_comparision} summarizes the differences between these and GAMA. %, comparing various aspects of the design.
\rev{CHARM \cite{zhuang:2023:fpga:charm, zhuang:2024:trets:charm_2} and AutoMM \cite{zhuang:2023:fpga:automm} accelerated GEMM on AIE using a resource conservative approach which enables efficient scaling but results in limited performance. CHARM and AutoMM show performance for the fp32 input and fp32 output (fp32-fp32) \cite{zhuang:2023:fpga:charm} and int8-int8 and int16-int16 \cite{zhuang:2023:fpga:automm} precisions, respectively. }
\gama maximizes resource utilization (such as AIEs and AIE-PL interfaces) to achieve maximum performance.
\maxeva \cite{taka:2023:fpt:maxeva} (fp32-fp32, int8-int32), on the other hand, maximizes GEMM performance while using significant resources on the chip. 
It uses two different kernels for matrix multiplication and reduction (addition).
This limits the overall efficiency of \maxeva to 80\% since only 80\% of the AIEs in the design can perform the matrix multiplication operation (20\% are used for addition). 
% It uses buffer data transfer method to communicate the partial sum.
On the other hand, \gama uses a single kernel that performs both matrix multiplication and reduction, similar to CHARM. \gama utilizes 94\% of the \aieml array for matrix multiplication.
It uses a cascade interface for partial sum transfer between AIEs, unlike \maxeva which uses buffer sharing between nearby engines for reduction.
AMA \cite{deng:2024:fpl:ama} uses a lower precision output (fp32-fp16, int8-int16) improving performance at the cost of potential accuracy loss. 
\maxeva and AMA provides AIE-only performance considering that input data is available in PL, unlike CHARM \cite{zhuang:2023:fpga:charm}, which provides a full implementation including PL and PS. 
% CHARM uses a cascade interface for partial sum transfer similar to us. 
% It uses packet switching techniques for PLIO reuse, unlike others who uses broadcasting. 
RSN-XNN \cite{wang:2025:arxiv:rsn_xnn} designs an overlay by proposing a reconfigurable stream network, leveraging the flexibility of FPGA.
% and heterogeneity of Versal. 

Some prior work has focused on the compilation for AIEs.
Vyasa \cite{chatarasi:2020:hpec:vyasa} extends the Halide \cite{ragan:2013:pldi:halide} DSL compiler to automatically generate code for AIEs, improving the programmability. 
ARIES \cite{zhuang:2025:fpga:aries} designs a code generator for AIE on Versal and AIE-ML on AMD Ryzen processors by leveraging the MLIR framework \cite{lattner:2020:axriv:mlir}.
\rev{The ARIES framework supports AIE-ML on Ryzen CPUs, 
%which are equipped with a limited quantity of AIEs. 
demonstrating performance results for ResNet instead of GEMM. 
Therefore, a direct comparison with ARIES is not feasible unless modifications are made to enable support for Versal devices.}
Wierse \cite{wierse:2023:ethz:evaluation} perform performance analysis of the AIE's communication interfaces. 
%H-GCN \cite{zhang:2022:fpl:hgcn} and Chen \cite{chen:2023:fpl:exploiting} deploy Graph Convolution Network and Graph Neural Network on Versal AIE1. 

Compared to prior work, GAMA is the first framework to show GEMM acceleration on AIE-ML. 
% Our work is the first to show GEMM performance on \aieml. 
% We draw inspiration from \maxeva and AMA for designing our kernels and connectivity while providing significant performance improvement over them. 
% \maxeva uses matrix multiplication kernels and a separate add kernel for partial sum reduction.
% \gama uses a single kernel that performs both matrix multiplication and reduction.
% This enables \gama to use all \aieml for matrix multiplication.
% \maxeva uses buffer data transfer method to communicate the partial sum, while \gama use the cascade stream interface.
% Both methods have different tradeoffs. 
% The overall efficiency of \maxeva is limited to 80\% since only 80\% of the AIEs in the design can perform the matrix multiplication operation (20\% are used for addition).
% However, the AIE cores that perform the addition help \maxeva relax buffer placements, reducing memory stalls. 
% \gama, on the other hand, utilizes 94\% of the \aieml array for matrix multiplication. 
% However, due to high density of matrix multiplication, the design becomes congested, resulting in memory stalls.
% Table \ref{tab:prior_work_comparision} further summarizes the differences between the previous work and ours, comparing various aspects of the design. 
\gama has the highest AIE memory utilization with custom buffer placement, a flexible pack size and a staggered kernel placement for efficient scaling. 
% The default compiler flags, and buffer optimization do not do a good job when it comes to buffer placement for a tight design. 
% our custom buffer placement avoid these memory stall.
% To summaries the tradeoff: Memory stall + Cascade stalls (Ours) vs AIE array utilization (MaxEVA).
% AIE-ML has a 512-bit cascade compared to VCK190 which is 384. This shows it’s cleaver to use the cascade instead of adder tree compared to MAXEVA

\section{Background}

%\subsection{Versal Architecture}
The AMD Versal AI Engine is a heterogeneous architecture that combines AI Engines, Programmable Logic (PL), and Processing System (PS).
The AI Engines are VLIW vector processors capable of executing multiple instructions per clock cycle.
Connected to the AI Engines is an FPGA fabric (or PL) consisting of Lookup Tables (LUTs), flip-flops (FFs), Digital Signal Processing slices (DSPs), Block RAMs (BRAMs), and Ultra RAMs (URAMs). 
The AI Engines communicate with the PL and DRAM through two interfaces: the PL interface input/output (PLIO) for data transfer between the PL and AI Engines, and NoC interface tiles for data transfer between the AI Engines and DRAM. 
The AI Engine array architecture provides us with Buffer (an AIE can directly access a neighboring AIE's memory), Cascade (direct AIE to AIE connectivity) and Via-switch connections to communicate from one AIE to another. 
The application processor can run a Linux operating system and can be used to control both PL and AI Engines.
In this work, we use the 2nd generation of the AI Engine called AIE-ML. 
% The next paragraph discusses the differences compared to AIE.
We further discuss the differences compared to AIE. 
% Next, we discuss the differences compared to AIE1.

\begin{figure}[h]
  \centering
  \includegraphics[width=0.8
  \linewidth]{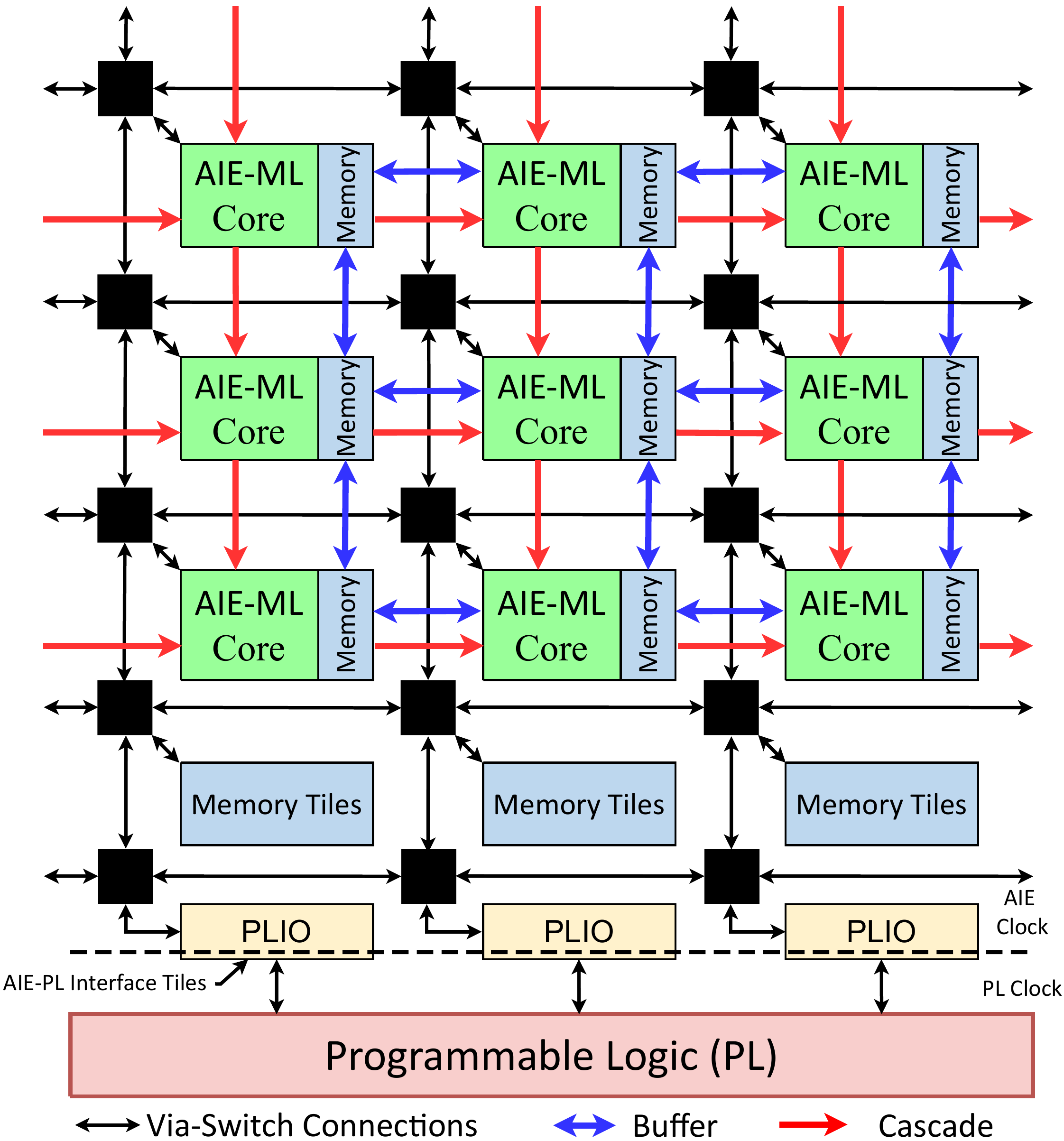}
  \caption{AMD Versal AIE-ML Architecture.}
  \label{fig:aie_ml_arch}
  \vspace{-5mm}
\end{figure}

%\subsection{AIE-ML Architecture}
Figure \ref{fig:aie_ml_arch} illustrates the Versal AIE-ML architecture. 
At the top are the AIE-ML cores that offer an int8 throughput of 256 MACs/cycle, which is double that of AIE. 
Additionally, AIE-ML features twice the memory of 64 KB compared to 32 KB in AIE. 
Although AIE-ML does not support fp32 computations, it incorporates support for bfloat16, contrasting AIE, which lacks bfloat16 but has fp32 support.
AIE-ML also includes a wider cascade bus of 512 bits compared to 384 bits in AIE. Although cascade connectivity remains unidirectional, AIE-ML now supports vertical cascading as well.
AIE-ML also provides us with addition memory storage in the form of memory tiles. 

% AIE to AIE communication interface: Cascade, Buffer and via-switch. 
% Versal ACAP architecture: AIE-ML. 
% Brief explanation of 
% AIE 
% PL 
% NOC 
% DDR 
% AIE-ML architecture 
% AIE-ML vs AIE differences 
% Compute 256 mac/cycle
% AIE – AIE connectivity: Cascade 512 bit 
% AIE memory 64 KB 

\section{GAMA Framework}
\label{sec:gama_framework}
% \vspace{-8mm}
In this section, we discuss the \gama framework in detail. 
%\gama accelerates GEMM workloads on \aieml maximizing the utilization of the \aieml array. 
We assume that the input/output data is available in the PL and use the PLIO interface for communication. 
We use AMD's Versal VE2802 device which has 304 engines in a grid of 8 rows x 38 columns. 
This device has 112 input PLIOs (PL to \aieml) and 84 output PLIOs (\aieml to PL). Each PLIO is configured to be 128 bit wide. 
\rev{
Matrix multiplication workload  
% In the context of matrix multiplication, matrix dimensions are
is specified with the notation M, K, and N. Where, matrix A has dimensions M $\times$ K, matrix B has dimensions K $\times$ N, and the resulting matrix has dimensions M $\times$ N.
}
% \vspace{-1mm}
% We use the same tiling scheme as \maxeva \cite{taka:2023:fpt:maxeva}. 

%This section is further organized as follows. We first discuss single \aieml workload size selection and performance optimizations followed by our implementation for the pack of \aieml[s]. Finally, we discuss our scaling strategy to scale to the complete \aieml array. 

\subsection{Single \aieml kernel design}\label{sec:gama_framework:single_aie}
% \vspace{-1in}
\noindent\textbf{Kernel size}:
Our kernels are designed using AMD APIs leveraging the generic (templated) matrix multiplication kernel from AMD's user API guide \cite{amd:2022:web:api_user_guide}, thereby ensuring their portability across a variety of Versal devices.
We support multiple sizes, i.e., the M, K, and N matrix dimensions, for a single kernel. 
% These dimensions should be divisible by 16, 8 and 16 respectively in order to support our kernel's vector read and write operations. 
Although our kernel code supports variety of M, K and N sizes, selecting the right size is crucial for high performance and \aieml kernel compute utilization. 
The kernel size selection is governed by the input/output precision, the \aieml memory limit and balance between compute and
PL-\aieml communication, which is defined by the compute-to-communication ratio \(\gamma\).
% We provide a systematic approach that maximizes \(\gamma\) and \aieml internal memory utilization. 
Equation \ref{eqa:theo_compute_cycles} calculates the theoretical kernel compute cycles based on the workload dimension and the peak throughput of the chip for a specific precision. 
Equations \ref{eqa:comm_a}, \ref{eqa:comm_b} and \ref{eqa:comm_c} calculate the cycles required to transfer the input matrix A and matrix B from PL to \aieml and the output matrix C from \aieml to PL, where $sizeof()$ depends on the precision used.
Each \aieml has two input PLIO and one output PLIO connections. 
This allows simultaneous pipelined read, compute, and write operations. 
Equation \ref{eqa:c_c_ratio} shows the calculation for \(\gamma\).
A ratio $< 1$ means the workload is PLIO bandwidth bound, while $> 1$ means it is compute bound. 
\allowdisplaybreaks
% \vspace{-7mm}
\begin{align}
\label{eqa:theo_compute_cycles}
& Compute\_cycles = \frac{M \times K \times N}{Peak\_MACs}  \\
&  \nonumber \\
\label{eqa:comm_a}
& Comm\_A=\ \frac{M\times K\times sizeof(input)}{(\frac{PLIO_{width}}{8})} \\
&  \nonumber \\
\label{eqa:comm_b}
& Comm\_B=\ \frac{K\times N\times sizeof(input)}{\left(\frac{PLIO_{width}}{8}\right)} \\
&  \nonumber \\
\label{eqa:comm_c}
& Comm\_C=\ \frac{M\times N\times sizeof(output)}{\frac{PLIO_{width}}{8}} \\
& \nonumber  \\
& \gamma = \frac{Compute\_cycles}{\begin{array}{c}
\max(Comm\_A, Comm\_B, Comm\_C)
\end{array}}
\label{eqa:c_c_ratio}
\end{align}

The internal memory of the AIE presents a constraint on the size of M, K and N as shown by Equation \ref{eqn:aie2_mem_constrain}. 
Since ping-pong buffering is used to overlap computation in \aieml and the \aieml-PL communication, the memory requirement is doubled.
\vspace{-5mm}
\begin{align}
& M \times K \times sizeof(input)  + \nonumber \\
& K \times N \times sizeof(input)   + \nonumber \\
& M \times N \times sizeof(output)   \times 2 \leq 64 \text{ KB} 
\label{eqn:aie2_mem_constrain}
\end{align}
\vspace{-2mm}

We perform an exhaustive search for M, K and N that satisfies the constraints in Equation \ref{eqn:aie2_mem_constrain} and has high \(\gamma\). 
% We assume a fixed bit width for PLIO and a fixed PL operating frequency for the search. 

\noindent\textbf{API sizing}:
AMD's API library \cite{amd:2022:web:api_user_guide} provides an API for matrix multiplication called MMUL, which performs blocked matrix multiplication C = A x B.
This API is defined as \texttt{\small{aie::mmul<M\_Elems, K\_Elems, N\_Elems, TypeA, TypeB, AccumTag>}} where \texttt{\small{M\_Elems, K\_Elems, N\_Elems}} define the size.
MMUL supports up to 9 sizes such as 4x4x4, 8x8x4 etc. 
%The next step is to identify the right MMUL API for the kernel. 
We use the sizes M, K, and N resulting from the constraints mentioned in the previous sub-section and then evaluate all available MMUL API sizes for the specified precision by performing a sweep.
We then select the MMUL API size that gives the best performance.
%No specific heuristic or logic is used to choose a particular MMUL API size.

\begin{figure}[h]
  \centering
  \includegraphics[width=0.8
  \linewidth]{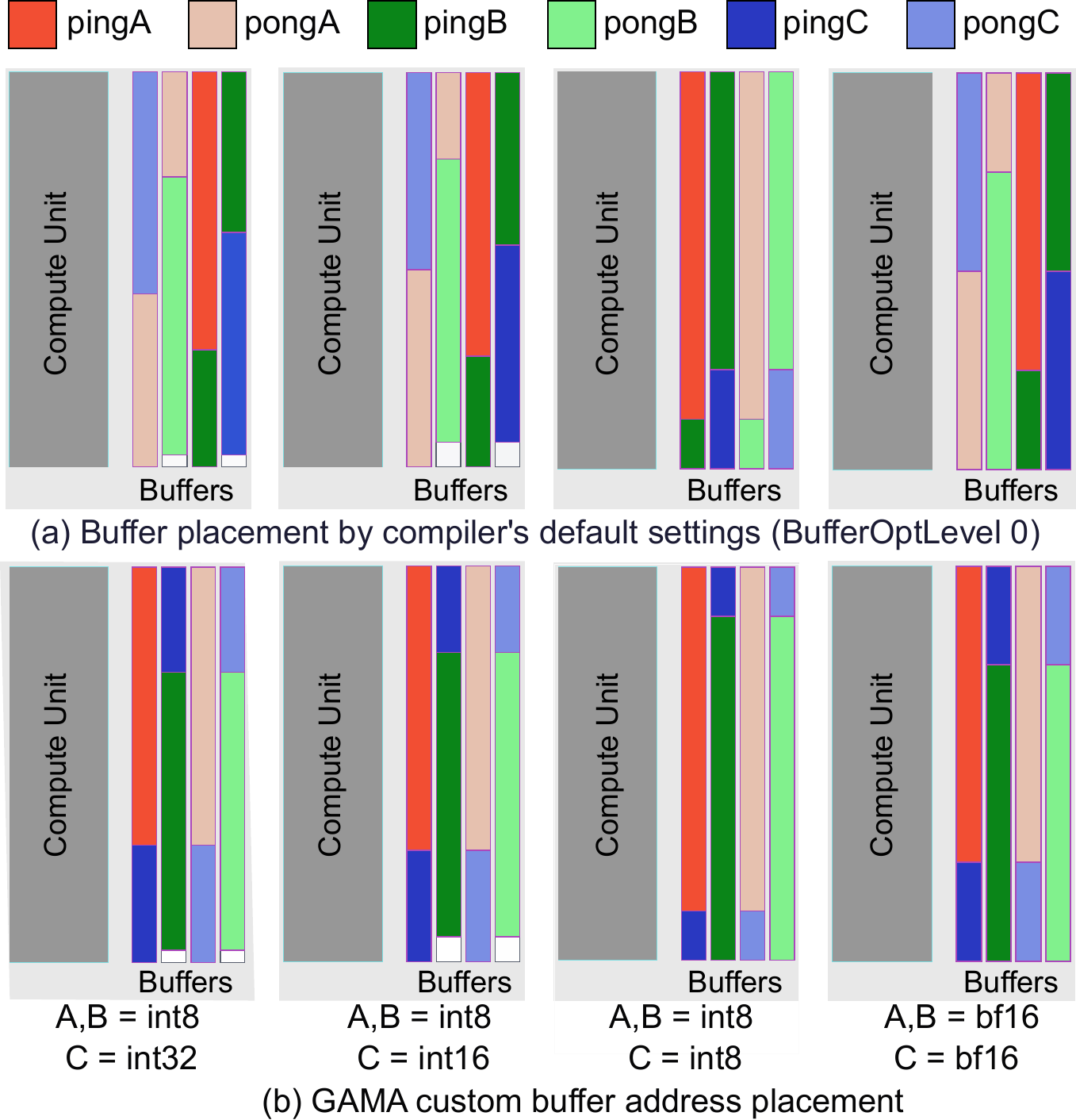}
  \caption{GAMA's custom buffer address placement within an AI engine minimizes stalls and improves performance.}
  \label{fig:buffer_placement}
\end{figure}

\noindent\textbf{Buffer placement}:
Each \aieml has 64 KB of AIE memory divided into 4 banks of 16 KB each.
Buffer placement or allocation assigns an address to a buffer inside the AIE memory's banks.
Optimal buffer placement in the AIE memory is crucial to prevent bank conflicts and subsequent memory stalls, thus improving kernel performance.
The AIE programming interface provides users with APIs to control buffer placement.
APIs can constrain the buffers to a certain AIE kernel location using  
\texttt{\small{location<buffer>(buf\_A) = location<kernel>(AIE\_kernel\_loc);}} or
can place the buffers at a specific address in the AIE's memory using 
\texttt{\small{location<buffer>(buf\_A) = \{address(row, col, <addr>),address(row, col, <addr>)\};}} 
providing a more finer control over the placement.
We refer to the former as ``buffer location placement" and the latter as ``buffer address placement". 
AMD also provides users with various compiler optimization flags to optimize buffer placement automatically.
% constraints, to enable fine-grained control over the design. 
Flags such as \bol
% and ``-xlopt"
\cite{amd:2022:web:kernel_graph_programming_guide} are used to optimize buffer allocation. 
% and loop performance inside the kernel, respectively. 
The \bol flag has 10 levels (0 to 9); increasing the level progressively increases optimizations made by the compiler to place the buffers to minimize memory stalls. 
These flags are effective without buffer placement constraints and with small buffers.

Every \aieml's kernel can directly access buffers located in its neighboring AIEs, enabling buffer placement in the neighboring AIEs. 
% Constraints can be used to restrict buffer allocation to a particular \aieml.
% Constraining buffers used by an AIE's kernel to that AIE makes it easy to scale the design to the \aieml array.
Constraining buffers using buffer location placement makes it easy to scale the design to the \aieml array.
With the buffer size that maximally utilizes the AIE memory and the above constraint, \bol 0 needs to be used for successful compilation 
(all other \bol levels give placement error).
But this leads to several memory stalls.

% We identify buffer locations using a custom buffer placement algorithm that maximizes \aieml memory utilization while minimizing stalls. 
We perform a custom buffer address placement using a simple algorithm that maximizes \aieml memory utilization while minimizing stalls.  
%Algorithm \ref{algo:buffer_placement} shows the pseudocode for the custom buffer placement. 
Our custom buffer address placement is guided by the set of rules that determine the buffer location based on the matrix type (input and output) and buffer category (Ping or Pong).
Our rules are as follows:
\textbf{(a)} Never assign ping and pong buffers of the same matrix to the same bank. 
\textbf{(b)} Never assign ping and pong buffers of the same matrix to the adjacent bank.
\textbf{(c)} Always assign the buffer of matrix A and B to a different bank.
\rev{Algorithm \ref{algo:buffer_placement} }encapsulates these rules and automates buffer address placement for various matrix sizes and precisions.
It performs a simple exhaustive search and generates buffer addresses that satisfy the rules mentioned above.
It takes the dimensions of matrix M, K, N, and the input and output precision as inputs and first calculates the size of buffers A, B, and C (lines 1-5).
If the sum of the sizes of the buffer exceeds 64 KB then it exits (Line 6).
A buffer list is initialized for all the buffers that need address allocation (Line 8). The order of the list is important. 
We iterate over all banks of to check if the corresponding buffer is suitable for allocation (Lines 10-32).
% We iterate over all banks over all buffers to check if the bank is suitable for allocation (Lines 9-31).
Banks have two spots for buffers, which means that max two buffers can fit in one bank.
Matrix A and B are only allocated when there are two spots available in the bank and the adjacent banks do not contain the corresponding double buffers (Lines 13-14).
If the constraint is satisfied, the buffer is placed in the bank, and a start address is allocated (lines 16-19). 
Matrix C can be placed as the second buffer in the banks that contain matrix A or B (Line 21).
The matix C buffer is placed and a start address is assigned to the buffer (Lines 22-26). 
The sum of both buffers can exceed the capacity (16 KB) of the bank, thus in this scenario the start address of the other buffer in the next bank is shifted by the offset (Lines 28-30).
Once finished, the algorithm returns the stating addresses of all six buffers. 

Figure \ref{fig:buffer_placement} (a) shows the compiler's default buffer placement with \bol 0 and (b) shows a result of our custom buffer address placement algorithm.
% for M=blah, N=blah, and K= blah, for precision = blah. 
%We show this for four different precisions. 
Our custom buffer address placement results in an average 12\% fewer cycles compared to automated buffer placement. We use the same algorithm for all buffer placements for the rest of the paper.

\begin{algorithm}
\caption{AIE buffer address placement algorithm }
\label{algo:buffer_placement}
\begin{algorithmic}[1]

\State \textbf{ip:} M, K, N, ip\_p, op\_p
\State \textbf{op:} buffer\_location
\State \textbf{Initialize:} aie\_mem.size = 65536 aie\_mem.bnks = 4

\State buf\_A, buf\_B \(\leftarrow\) \(M \times K \times ip\_p\), \(K \times N \times ip\_p\)
\State buf\_C \(\leftarrow\) \(M \times N \times op\_p\)
% Check for memory overflow
\If{\Call{chk\_ofl}{M, K, N, ip\_p, op\_p, aie\_mem.size}}
    \State \textbf{exit}
\EndIf

% Create buffer list
\State buf\_list[:] \(\leftarrow\) \small{ping\_A,pong\_A,ping\_B,pong\_B,ping\_C,pong\_C}
\State bnks \(\leftarrow\) aie\_mem.bnks
\For{buf in buf\_list}
    \For{b in range(bnks)}        
        % Case 1: Buffers for matrices A or B
        \If{ \(\text{buf} \in (Mat_A or Mat_B)\)}
            % Skip if adjacent ping-pong buffers exist or bank is full
            \If{\Call{is\_adjacent}{buf,b,aie\_mem} OR 
                b.free\_spots $\leq$ 1}
                \State \textbf{continue}
            % Otherwise allocate buffer to bank
            \Else
                % Allocate buffer to bank and update free space/spots
                % Set buffer start address based on bank start address
                % Break loop for current buffer
                \State bnks[b].buffer \(\leftarrow\) buf
                \State bnks[b].free\_space $-=$ buf.size
                \State bnks[b].free\_spots $--$
                \State buf.start\_addr \(\leftarrow\) bnks[b].start\_addr
            \EndIf

        % Case 2: Buffers for matrix C (op)
        \ElsIf{ \(\text{buf} \in (Mat_C)\)}
            \If {bnks[b].free\_spot $>$ 0}
                \State bnks[b].buffer \(\leftarrow\) buf
                \State bnks[b].free\_space $-=$ buf.size
                    \If {bnks[b].free\_spots $=$ 2}
                        \State buf.s\_addr \(\leftarrow\) bnks[b].s\_addr
                    \EndIf
                    \If {bnks[b].free\_spots $=$ 1}
                        \State buf.s\_addr \(\leftarrow\) bnks[b].s\_addr + bnks[b].buffer[0].size 
                    \EndIf
                    \If {bnks[b].free\_space $<$ 0}
                        \State offset \(\leftarrow\) (bnks[b].buffer[0].size + bnks[b].buffer[1].size) - bnks[b].size
                        \State bnks[b+1].buffer.s\_addr \(\leftarrow\) bnks[b+1].s\_addr + offset
                        \State bnks[b].free\_spots$--$
                    \EndIf
                    \State \textbf{break}
                \EndIf  
        \EndIf
    \EndFor
\EndFor

\State \textbf{return} buf\_list.addr
\end{algorithmic}
\end{algorithm}

\vspace{-5mm}

\subsection{Chaining the AIEs together} \label{sec:gama_frame:pack}
\vspace{-4mm}
\begin{figure}[h]
  \centering
  \includegraphics[width=0.6
  \linewidth]{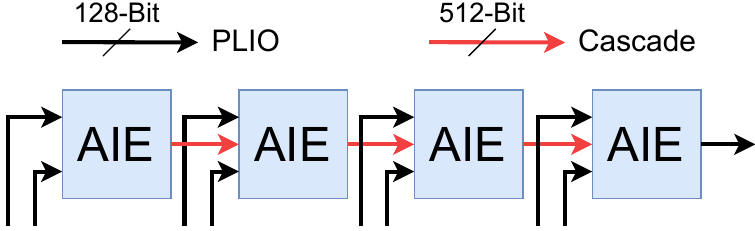}
  \vspace{-2mm}
  \caption{Pack of 4 AIEs computing a larger GEMM by using the cascade interface for reductions. Final workload size for the pack in terms of M, K and N of a single AIE kernel $\rightarrow$ M=M, K=4*K and N=N}
  \label{fig:aie_pack_4}
  \vspace{-3mm}
\end{figure}

\begin{figure}[t]
  \centering
  \includegraphics[width=0.9
  \linewidth]{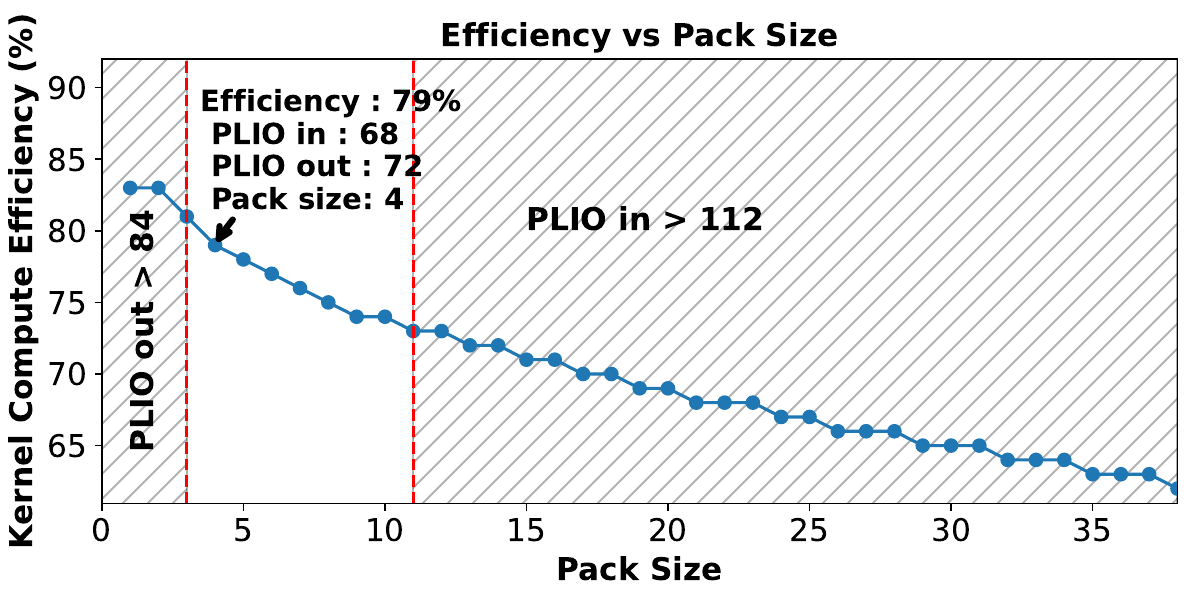}
  \vspace{-5mm}
  \caption{Efficiency drops as pack size increases. Pack size of 4 has the maximum efficiency among the sizes that can scale well to the whole AIE array (unhatched portion).}
  \label{fig:eff_vs_pack_size}
  \vspace{-4mm}
\end{figure}

\begin{figure}[h]
  \centering
  \includegraphics[width=0.7
  \linewidth]{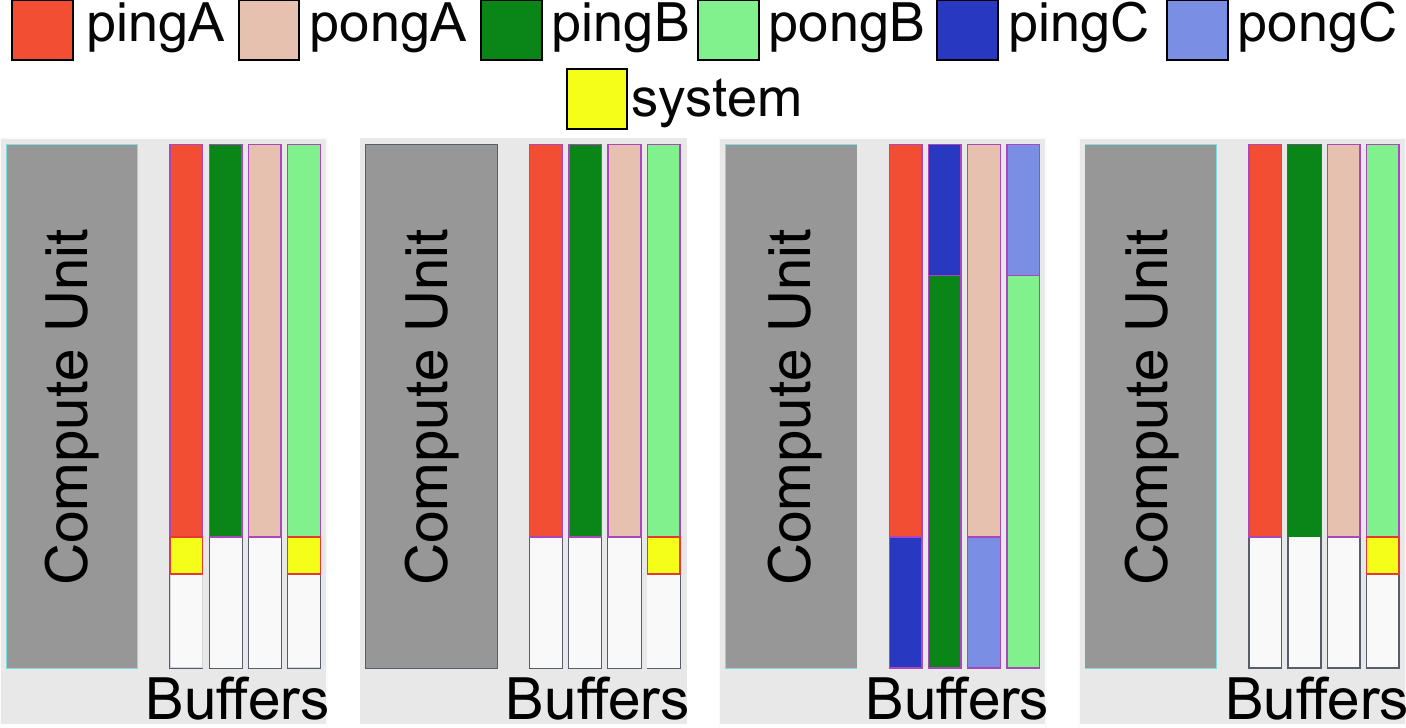}
  \caption{Buffer address placement for a pack of four AIEs that perform a GEMM operation together. Note the output buffers are placed in the 3rd AIE.}
  \label{fig:aie_pack_4_buffer}
\vspace{-6mm}
\end{figure}

\noindent\textbf{AIE to AIE connectivity}:
\aieml[s] are connected in series, as shown in Figure \ref{fig:aie_pack_4}, to form a pack, allowing larger GEMMs to be implemented with increased parallelism and reuse. 
There are two PLIOs for input to each \aieml in the pack and one PLIO for output from the pack. 
Reductions are performed inside the pack, and only final results are written back to the PL. 
This reduces the output PLIOs in the design, since only the last \aieml has to write data back to the PL. 
Cascade, via-switch connection, and buffer are the various methods for partial sum communication from one \aieml to the next \aieml. 
%AMA \cite{deng:2024:fpl:ama}, 
CHARM \cite{zhuang:2023:fpga:charm}, ARIES \cite{zhuang:2025:fpga:aries} and RSN-XNN \cite{wang:2025:arxiv:rsn_xnn} use a cascade interface, while \maxeva \cite{taka:2023:fpt:maxeva} and AMA \cite{deng:2024:fpl:ama}
use buffer. 
Buffer communication does not stream data from one AIE to another; it uses buffers in the neighboring AIE's memory.
These buffers consume memory.
% Buffer communication uses extra memory as the double buffers are required for overlap. 
Via-switch and cascade connections stream data from one AIE to another avoiding this memory overhead.
Via-switch interface is only 32-bit and using it increases communication latency significantly. 
\aieml architecture has a 512-bit wide cascade interface compared to AIE's 384-bit. 
The cascade with the larger bit width and no memory overhead becomes the best candidate for AIE to AIE communication.
% Thus, using a cascade preserves \aieml memory while also enhancing performance.
We use cascade interface for all partial sum communication. 
Cascade connections make the pack run in a data-flow fashion. 
\begin{figure*}[t]
  \centering
  \includegraphics[width=0.8
  \linewidth]{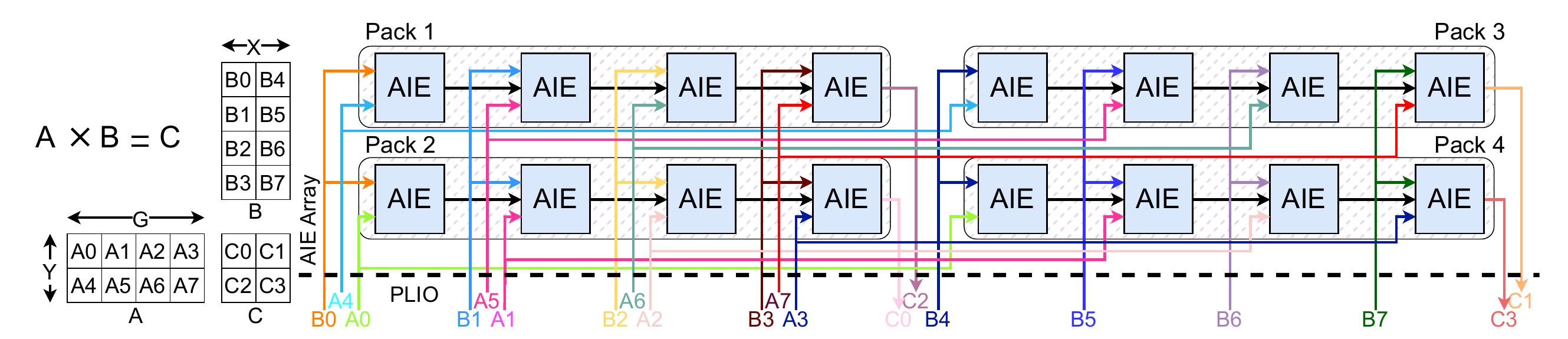}
  \vspace{-3mm}
  \caption{Our scaling and mapping strategy shown for hyperparameter settings of Y=2, G=4, X=2. Final workload size in terms of M, K and N of a single AIE kernel $\rightarrow$ M=Y*M, K=G*K and N=X*N}
  \label{fig:aie2_mapping_2x4x2}
  \vspace{-6mm}
\end{figure*}
\noindent\textbf{Different kernel types}:
The pack includes three distinct kernels.
The first kernel (used in the leftmost \aieml in the pack) takes Matrix A and Matrix B as inputs from the PLIO and output's partial sum through cascade interface.
The second kernel (used in the middle \aieml[s] in the pack) has two input PLIO ports, one input cascade port, and one output cascade port. 
%This kernel is replicated as required when the pack size is greater than 3. 
The third kernel (used in the last \aieml in the pack) receives two inputs from PLIO, one input from the cascade port, and one output PLIO port. % to write back results to the PL.

\noindent\textbf{Pack size}:
Due to the flexibility of the design, any number of \aieml[s] can be chained as long as PLIO resources are available to support the design. 
We calculate the kernel compute efficiency (KCE) for the pack to gauge the effect of cascade on the design.
The average the kernel compute cycles across all the AIE's in the pack is used to calculate the KCE.
Figure \ref{fig:eff_vs_pack_size} shows how the kernel compute efficiency varies as we change the pack size from 2 to 38.
The max pack size is 38 in our case because VE2802 (the device we use) only has 38 columns in the AIE grid, and we use horizontal cascade connections.
%Other devices with more columns can have a larger pack size.
Designs for all of these pack sizes compile and run, but not all of them scale well. 
The hatched portion in the figure shows the pack sizes that do not scale to the complete array, as they run out of input PLIO or output PLIO resources (PLIO requirement increases with pack size).
Small packs (2-3) are bound by output PLIOs, and larger packs (11-38) are bound by input PLIOs.
Thus, only pack sizes of 3-10 are scalable. 
Although the cascade interface does not have memory overhead and has a large bit width, it incurs cascade stalls. 
Cascade stalls occur when the producer kernel tries to write data elements that exceed the bit width of the cascade interface. 
These stalls accumulate as we chain more \aieml[s], leading to a reduction in kernel compute efficiency as the pack size increases.
%Figure \ref{fig:eff_vs_pack_size} shows the drop in kernel compute efficiency as we increase the pack size. 
We select a pack size of 4 as it has the highest kernel compute efficiency and is scalable to the \aieml array. 
% We employ a similar buffer placement strategy for the pack, ensuring that all buffers are allocated within the same pack to maintain scalability. 

AIEs connected in a pack perform a GEMM operation with matrix sizes larger than the matrix sizes for a single AIE.
% The pack affects the native K dimension of the matrix. 
For example, if the kernel sizes are M=48, K=240 and N=48 then the native size of the pack becomes M=48, K=960 and N=48.
% Add lines here to say what is the native size of the GEMM performed in a pack of 4 if the dimensions of the single kernel are M,K,N
% The MKN kernel sizes derived in Section \ref{} are utilized consistently throughout this process.
% We restrict the group to 4 AIEs, This groups scales pretty well across the complete array of AIEs. (Needs more explanation here)

\noindent\textbf{Buffer placement}:
We use cascade interface to transfer the data from one AIE to the next AIE in a pack as seen in the previous section. 
Cascade avoids the need for output buffers. 
Thus, only the last AIE in the pack generates results for the overall GEMM operation performed by the pack.
%, all the other AIEs only have input buffers.
Figure \ref{fig:aie_pack_4_buffer} shows the placement of the input and output buffers for a pack size of four. 
In the pack of four, the 4th AIE generates the output, but we place the output buffers of the 4th AIE into the 3rd AIE.
Doing this further avoids memory stalls. 
The three AIEs (1st, 2nd, and 4th) have only 4 input buffers each, so they do not need custom buffer address placement.
% due to the relaxed constraints. 
However, the 3rd AIE will have all six buffers (4 input buffers and 2 output buffers) and therefore needs the custom buffer address placement from Section \ref{sec:gama_framework:single_aie}. 
All buffers are contained within the pack, simplifying scaling to the AIE array.
% We use our algorithm to find the buffer locations for individual \aieml in a pack. 
% OUTPUT BUFFER PLACEMENT in the 3rd AIE in the pack
\vspace{-2mm}
\subsection{Scaling to the complete \aieml array}
% Table generated by Excel2LaTeX from sheet 'Results section '
\vspace{-1mm}
\setcounter{table}{2}
\begin{table*}[t]
  \centering
  \begin{threeparttable}
        \caption{Single \aieml Kernel Compute Cycles 
        \label{tab:sa_buff_plmt_res_mmul}%
        \textbf{(KCC)} and Kernel Compute Efficiency \textbf{(KCE)} for different buffer placement strategies}
        \begin{tabular}{|c|c|c|c|c|c|c|c|c|c|c|c|c|}
        \hline
        \multicolumn{5}{|c|}{\textbf{Configuration} } & \multicolumn{1}{c|}{\multirow{2}[1]{3em}{\textbf{KCC (Theor- etical})}} & \multicolumn{2}{p{10em}@{}|}{\textbf{Unconstrained buff \tnote{a,\$} \textcolor{red} }} & \multicolumn{2}{p{12.5em}|}{\textbf{Buffer location placement \tnote{b,\$}}} & \multicolumn{3}{c|}{\textbf{Buffer address placement \tnote{c,\$}}} \bigstrut \\
        % \hline
        \cline{1-5}\cline{7-13} \multicolumn{1}{|p{3em}|}{\textbf{Precision (ip-op)}}  & \multicolumn{1}{p{1em}|}{\textbf{M}} & \multicolumn{1}{p{1em}|}{\textbf{K}} & \multicolumn{1}{p{1em}|}{\textbf{N}} 
        & \multicolumn{1}{p{3em}|}{\textbf{MMUL API}} &  & \multicolumn{1}{p{5em}|}{\textbf{KCC (Measured)}} & \multicolumn{1}{p{3em}|}{\textbf{KCE (\%)}} & \multicolumn{1}{p{6em}|}{\textbf{KCC (Measured)}} & \multicolumn{1}{c|}{\textbf{KCE (\%)}} & \multicolumn{1}{p{5em}|}{\textbf{KCC (Measured)}} & \multicolumn{1}{p{2em}|}{\textbf{KCE (\%)}} & \multicolumn{1}{p{5em}|}{\textbf{\%Perf recovered}} \bigstrut\\
        \hline
        \textbf{int8-int32} & 48 & 240   & 48 & 4x8x8   & 2160  & 2426  & 89\% & 3076  & 70\% & 2590  & 83\% & 13\% \bigstrut\\
        \hline
        \textbf{int8-int16} & 64    & 184   & 64  & 4x8x8    & 2944  & 3141  & 94\% & 3923  & 75\% & 3345  & 88\% & 13\% \bigstrut\\
        \hline
        \textbf{int8-int8} & 64    & 224   & 64    & 4x8x8  & 3584  & 3686  & 97\% & 4340  & 83\% & 3831  & 94\% & 11\% \bigstrut\\
        \hline
        \textbf{bf16-bf16} & 64 & 96    & 64  & 8x8x4    & 3072  & 3135  & 98\% & 3598  & 85\% & 3255  & 94\% & 9\% \bigstrut[t]\\
        \hline
        \end{tabular}%
        \begin{tablenotes}
            \item[a] Unconstrained buffer placement with \bol 9. 
            \item[b] Buffers constrained to same AIE using buffer location placement and \bol 0.
            \item[c] Buffers constrained to same AIE using buffer address placement and \bol 0.
            \item[\$] Per-AIE memory utilization is shown in Table \ref{tab:sa_c_c_mem_util}
        \end{tablenotes}
    \end{threeparttable}
  \vspace{-7mm}
\end{table*}%

To maximize performance, the design of a pack is scaled across the \aieml array by replicating the pack.
% We use broadcasting feature in the PLIO interface tile to reuse the PLIOs. 
% while ensuring efficient data reuse and minimal bottlenecks. 
%This subsection outlines the scaling methodology that includes the use of hyperparameters, constraints, and custom placement.
To improve data reuse, we leverage the PLIO broadcast mechanism. 
This minimizes redundant data transfers and optimizes overall performance.
Broadcasting also saves a lot of PLIOs by enabling reuse, which 
is crucial since PLIO interfaces are limited. 

\noindent\textbf{Scaling hyperparameters and constraints}:
The scaling in our design is governed by three hyperparameters.
\textbf{Y}: Replicates the pack along the Y-axis of the \aieml array.
\textbf{G}: Defines the size of each pack.
\textbf{X}: Replicates the pack along the X-axis of the \aieml array.
%These parameters allow for flexible scaling while adhering to architectural constraints.
Figure \ref{fig:aie2_mapping_2x4x2} shows the scaling for the case (Y = 2,G = 4,X = 2). 
PLIOs are shown using different colors that show broadcast and reuse. 
Matrix A and matrix B are assigned to 16 \aieml as shown in Figure \ref{fig:aie2_mapping_2x4x2}.
% The same mapping can be scaled to the complete array. 

Scaling to the complete \aieml array must satisfy two main constraints: (1) 
The first constraint, shown in Equation \ref{eqn:scaling_constraints_device}, ensures compatibility with the geometry of the VE2802 device.
(2) The second constraint in Equation \ref{eqn:scaling_constraints_aie_plio} ensures the total PLIOs and \aieml[s] do not surpass available resources.

\vspace{-5mm}
\begin{align}
\nonumber \
\label{eqn:scaling_constraints_device} \
&Y \le \# of\ rows\ in\ the\ device (8) \\
&G\times X \le \# of\ columns\ in\ the\ device\ (38)\\ 
\nonumber \
\end{align}
\vspace{-15mm}
\begin{align}
\nonumber \
& Y\times G\times X\le AIE\_cores \\
\nonumber \
& Y\times G+G\times X\le PLIO\_in \\
& Y\times X\le PLIO\_out\
\label{eqn:scaling_constraints_aie_plio}
\end{align}
\vspace{-5mm}

\begin{figure}[t]
  \centering
  \includegraphics[width=0.95
  \linewidth]{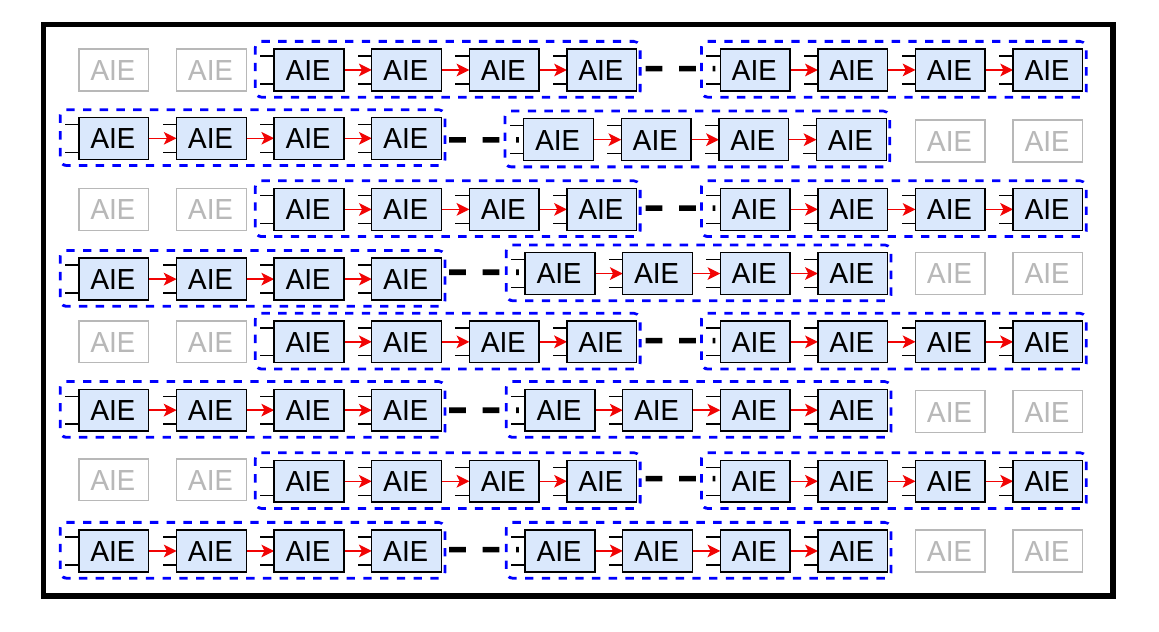}
  \vspace{-5mm}
  \caption{A staggered placement is used while replicating pack of kernels across the whole array to avoid routing congestion.}
  \label{fig:aie_array_scaling}
\vspace{-6mm}
\end{figure}

\noindent\textbf{Kernel placement}:
Scaling the design to the entire array by satisfying the constraints stated above leads to compilation failure due to PLIO routing congestion. 
This happens because the third AIE utilizes three distinct PLIOs (two read, one write), leading to congestion in the switch network's vertical lane when extended across all rows. 
To avoid this congestion, \gama uses a zigzag or skewed kernel placement strategy across the array, as shown in Figure \ref{fig:aie_array_scaling}. 
We skew the kernel placement by two since skewing by one also leads to congestion, and skewing by three reduces the AIE array utilization.
The first two \aieml[s] in each alternate rows of \aieml[s] are not used.
The pattern alternates the third AIE's row position, easing vertical switch lane pressure.
This placement successfully avoids congestion, leading to a scalable and compilable design.
% This manual kernel placement
% is performed at the full array level, ensuring optimal resource utilization and scalability.

Note that this manual kernel placement and the custom buffer address placement from the previous subsection reduces the compiler's efforts, lowering the compilation time by 6x. 
%Furthermore, our custom manual buffer placement significantly reduces compile time by eliminating the need for automated buffer allocation by the compiler. This approach also ensures seamless scalability for the entire array.

\section{Evaluation Results}
\begin{comment}
% In this section, we first provide details about the experimental setup, then show performance on a single \aieml and then for a pack of \aieml[s] and finally on the complete array of \aieml. 
% We provide analysis and results for two main precisions int8 and bf16. 
In this section, we begin by describing the experimental setup. 
We then present performance results for a single AI-Engine, followed by a pack of AI-Engines, and finally for the entire array. 
Additionally, we provide results for two key precisions: int8 and bf16.
% We show the advantage of \gama's custom buffer placement compared to the default buffer placement 
\end{comment}

\subsection{Experimental setup}

\setcounter{table}{1}
% Table generated by Excel2LaTeX from sheet 'Results section '
\begin{table}[t]
  \centering
  \begin{threeparttable}
  \caption{Single \aieml compute to PLIO communication ratio (\(\gamma\)) and \aieml memory utilization}
  \label{tab:sa_c_c_mem_util}%
    \begin{tabular}{|@{}p{5em}@{}|c|c|c|p{1.2em}|p{3em}|p{4.2em}@{}|p{4em}|}
    \hline
    {\textbf{Precision (ip-op)}} & {{\textbf{M}}} & {{\textbf{K}}} & {{\textbf{N}}} & {\textbf{{\(\gamma\) }}} & \textbf{AIE Mem Usage\tnote{a}} & \textbf{AIE Mem UCB\tnote{b} (\%)}& \textbf{AIE Mem CB\tnote{c} (\%)} \bigstrut \\
    \hline
    \textbf{int8-int32} & 48    & 240   & 48    & 0.72  & 64512 & 63\% & 98\% \bigstrut \\
    \hline
    \textbf{int8-int16} & 64    & 184   & 64    & 0.96  & 63488 & 25\% & 97\% \bigstrut \\
    \hline
   \textbf{int8-int8} & 64    & 224   & 64    & 0.96  & 65536 & 12\% & 100\% \bigstrut \\
    \hline
    \textbf{bf16-bf16} & 64    & 96    & 64    & 0.96  & 65536 & 25\% & 100\% \bigstrut \\
    \hline
    \end{tabular}%
    \begin{tablenotes}
    \item[a] AIE Memory usage for the M, K and N size matrix
    \item[b] Per-AIE Memory utilization for unconstrained buffer placement
    \item[c] Per-AIE Memory utilization for constrained buffer location and address placement
    \end{tablenotes}
    \end{threeparttable}
\vspace{-8mm}
\end{table}%

Our experiments are carried out using the VE2802 Versal AIE-ML device on the VEK280 board using Vitis version 2024.1. 
We utilize the \texttt{aiesimulator} tool offered by AMD to simulate our designs and measure 
%\red{power} and 
performance.
% All our experiments use 1.25 GHz as the \aieml operating frequency and 300 MHz as the PL frequency. We used a 128-bit PLIO configuration for both input and output PLIOs throughout the experiments.
AIEs operate at a frequency of 1.25 GHz, while the PL frequency is set to 300 MHz. A 128-bit PLIO configuration is used for both input and output throughout the experiments.
AIE-to-AIE communication is done using the 512-bit cascade interface.
\vspace{-2mm}
\subsection{Performance of single AIE kernels}
\setcounter{table}{3}
% Table generated by Excel2LaTeX from sheet 'Results section '
\begin{table*}[h]
  \centering
  \begin{threeparttable}
  \caption{ Buffer placement's effect on performance for a pack of 4 \aieml[s]}
  \label{tab:pack_buff_plmt_res}%
    \begin{tabular}{|c|c|c|c|c|c|c|c|c|c|c|c|c|}
    \hline
    \multicolumn{4}{|c|}{\textbf{Configuration} } &  \multicolumn{3}{c|}{\textbf{Unconstrained buff \tnote{a}}} & \multicolumn{2}{c|}{\textbf{Buffer location placement \tnote{b}}} & \multicolumn{3}{c|}{\textbf{Buffer address placement \tnote{c}}} \bigstrut \\
    \hline
    \multicolumn{1}{|p{3em}|}{\textbf{Precision (ip-op)}}  & \multicolumn{1}{p{1em}|}{\textbf{M}} & \multicolumn{1}{p{1em}|}{\textbf{K}} & \multicolumn{1}{p{1em}|}{\textbf{N}} & \multicolumn{1}{p{1em}|}{\textbf{KCC\tnote{*}}} & \multicolumn{1}{p{5em}|}{\textbf{KCE (\%)}} & \multicolumn{1}{p{5em}|}{\textbf{\% Cascade stalls}} & \multicolumn{1}{c|}{\textbf{KCC\tnote{*}}}  &\multicolumn{1}{c|}{\textbf{KCE (\%)}} & \multicolumn{1}{c|}{\textbf{KCC\tnote{*}}} & \multicolumn{1}{c|}{\textbf{KCE (\%)}} & \multicolumn{1}{p{6.835em}|}{\textbf{\% performance recovered }} \bigstrut\\
    \hline
    \textbf{int8-int32} & 48    & 960   & 48    & 2665  & 81\%  & 9\%   & 3198  & 68\%  & 2711  & 80\%  & 12\% \bigstrut\\
    \hline
    \textbf{int8-int16} & 64    & 736   & 64    & 3326  & 89\%  & 6\%   & 4126  & 71\%  & 3419  & 86\%  & 15\% \bigstrut\\
    \hline
    \textbf{int8-int8} & 64    & 896   & 64    & 3980  & 90\%  & 7\%   & 4273  & 84\%  & 4009  & 89\%  & 6\%  \bigstrut\\
    \hline
    \textbf{bf16-bf16} & 64    & 384    & 64    & 3361  & 91\%  & 7\%   & 4340  & 71\%  & 3404  & 90\%  & 19\% \bigstrut\\
    \hline
    \end{tabular}%
    \begin{tablenotes}
    \item[a] Unconstrained buffer placement with \bol 9.
    \item[b] Buffers constrained to same AIE using buffer location placement and \bol 0.
    \item[c] Buffers constrained to same AIE using our buffer address placement with \bol 0.
    \item[*] All KCC values are measured cycles averaged across 4 AIEs. (M, K, N) denotes the final size of the GEMM operation on the pack of 4 AIEs  
    \end{tablenotes}
    \end{threeparttable}
  \vspace{-6mm}
\end{table*}%

% We have designed our matrix multiplication kernels using AMD's API \cite{}. We used the MMUL API to perform blocked matrix multiplication. 
At the single AIE level, the two most important aspects are the MMUL API size and the single AIE kernel size (MxKxN).
% Initially, focus on choosing the appropriate \aieml workload size, followed by selecting the suitable MMUL API to achieve optimal performance.
As discussed in Section \ref{sec:gama_framework:single_aie}, the size of a single AIE kernel depends on factors such as the compute-communication ratio \(\gamma\), AIE memory usage and input/output precision. 

We do an exhaustive search for M, K and N sizes, and then
% and then calculate \(\gamma\).
% a low value of \(\gamma\) indicates that a workload is communication bound whereas a high value indicates that the workload is compute bound.
% Different precision results in different sizes for M,K and N. 
select the ones with the highest value of \(\gamma\) while satisfying the memory requirements stated in Equation \ref{eqn:aie2_mem_constrain}.
% ensuring that internal memory usage stays below 64 KB (the AIE memory size in the architecture). 
Different precisions result in different sizes for M,K and N.
% The values of M, K, and N also vary depending on the precision used.

Table \ref{tab:sa_c_c_mem_util} presents the results of this exhaustive search. For each precision, we show the kernel size (M, K, and N) along with the compute-to-communication ratio \(\gamma\) and the internal memory usage. 
Our constrained buffer address placement achieves 98\%, 97\%, 100\%, and 100\% AIE memory utilization for int8-int32 (ip-op), int8-int16, int8-int8, and bf16-bf16 precisions, respectively.
% For int8-int32 (ip-op), int8-int16, int8-int8, and bf16-bf16 precision, the AIE memory utilization is observed to be 98\%, 97\%, 100\%, and 100\%, respectively. 
\gama achieves the highest memory utilization compared to all other prior work \cite{taka:2023:fpt:maxeva,zhuang:2023:fpga:charm,deng:2024:fpl:ama,wang:2025:arxiv:rsn_xnn}.
All precisions except int8-int32 achieve a 0.96 compute-to-communication ratio, as int8-int32's larger output size increases communication latency.

To calculate the efficiency of the kernel, we measure the kernel compute cycles (KCC) and then take the ratio of theoretical KCC to measured KCC and call it kernel compute efficiency (KCE).
We use KCE as a metric to find the best MMUL API size. 
We sweep all MMUL sizes to find the optimal M, K, and N sizes, as noted in Section \ref{sec:gama_framework:single_aie}.
Table \ref{tab:sa_buff_plmt_res_mmul} shows the KCE for the best respective MMUL size for each precision. 
%We perform this experiment without placing any constraints on the buffer placement and set the \bol to 9 to avoid memory stalls. 
% We perform a sweep across all MMUL API sizes and select the one with the best performance (lowest execution cycles). 
% We measure Kernel Compute Efficiency (KCE) as the ratio of the theoretical kernel compute cycles (KCC) to the measured KCC.
% Table \ref{tab:sa_buff_plmt_res_mmul} shows the kernel computation efficiency (KCE) for the best respective MMUL size.
% We achieve on average 94\% KCE for all precisions. 
% We did not impose any constraints on buffer positioning and utilized \bol set to 9.
% As discussed in \ref{sec:gama_framework} this value of \bol is not scalable.
% (We omit adding the sweep results to the paper due to page constrains, we only mention the MMULs with best performance).
% Although \bol of 8 shows the best performance for single \aieml designs, having no constraints on the buffers makes it hard for the compiler to place a larger design efficiently on the array. 
% In our experiments for the single \aieml workload sizes we selected, the compiler fails to place the buffer with \bol of 8 inside a single \aieml. 
% We tried all possible levels from 0 to 9, but only level 0 works, rest of the levels result in placer giving error that it cannot place the design. 
% With a \bol of 0, this placement is possible, and the design is compiled, but this level has poor performance, resulting in a huge amount of memory stalls.
The table %Table \ref{tab:sa_buff_plmt_res_mmul} 
compares the KCC and KCE for three scenarios: unconstrained buffers (\bol 9), buffers constrained to the same AIE using buffer location placement (\bol 0), and then buffer constrained to the same AIE using our buffer address placement. 
% compares the KCC and KCE for three scenarios: unconstrained buffers (\bol 9), constrained buffers (\bol 0), and our custom buffer placement.
Unconstrained single AIE buffers show the best performance with the highest average KCE of 94\%.
However, as mentioned earlier, this is not scalable because the buffers can be allocated in neighboring AIEs by the compiler.
It significantly reduces AIE memory utilization as shown in Table \ref{tab:sa_c_c_mem_util}.
We use this best-case performance as a baseline for further comparison.
% We consider a single unconstrained AIE as the baseline for further comparison with the highest average KCE of 94\%. 
Buffers constrained to the same AIE using buffer location placement and \bol 0 shows an average 16\% drop in KCE. 
% Using constrained buffers, we observe a 16\% drop in KCE. 
The observed drop occurs due to the random fragmentation of buffers across the AIE banks.
However, buffers constrained to the same AIE using our custom buffer address placement recover on average 12\% of this loss, bringing the average KCE to 90\%, while maximizing AIE memory utilization and ensuring scalability.
% However, our custom buffer placement on average recovers nearly 70\% of this loss, bringing the average KCE to 90\%.
% While our custom 
% Thus, we use our custom placement strategy to carefully place the individual buffers and their double buffers inside a single \aieml. 
% Our custom placement results in a compilable design and shows fewer memory stalls, improving the performance of the design.
% Table \ref{tab:sa_buff_plmt_res} illustrates the performance drop caused by setting \bol to 0 and highlights the performance recovered as a result of our custom buffer placement method.
% The single AIE sizes and MMUL size used here are applied throughout the rest of the results section.

\setcounter{table}{4}
\begin{table}[t]
  \centering
  \begin{threeparttable}
  \caption{Performance scaled to the whole AIE array. }
    \label{tab:aie_array_final_perf}%
    \begin{tabular}{|c|c|c|c|c|c|c|c|c|}
    \hline
    \multicolumn{1}{|p{4em}|}{\textbf{Precision (ip-op)}}  & \multicolumn{1}{p{1em}|}{\textbf{M}} & \multicolumn{1}{p{1em}|}{\textbf{K}} & \multicolumn{1}{p{1em}|}{\textbf{N}} &\multicolumn{1}{p{7em}|}{\textbf{Throughput (TOPs/TBFLOPs)}}  &\textbf{TE}\bigstrut\\
    \hline
    \textbf{int8-int32}   & 384     & 960     & 432    & 133 &69\%   \bigstrut\\
    \hline
    \textbf{int8-int16}   & 512     & 736     & 576    & 159 &82\%    \bigstrut\\
    \hline
    \textbf{int8-int8}   & 512     & 896     & 576   & 165 &85\%    \bigstrut\\
    \hline
    \textbf{bf16-bf16}   & 512     & 384     & 576   & 83 &86\%    \bigstrut\\
    \hline
    \end{tabular}%
    \begin{tablenotes}
    \item[] (M, K, N) denotes the final size of the GEMM.% operation mapped to the AIE array.
    \item[] AIEs = 288 (94.7\%), Memory Banks = 2304 (94.7\%), PLIOs In = 68 (60.7\%), PLIOs out = 72 (85.7\%), Y=8, G=4, X=9.
    \end{tablenotes}
    \end{threeparttable}
\vspace{-4mm}
\end{table}%

\setcounter{table}{5}
% Table generated by Excel2LaTeX from sheet 'Results section '
\begin{table}[h]
  \centering
  \caption{Relative performance comparison with prior work, comparing the throughput efficiency (TE) 
  }
  % (i.e. ratio of achieved throughput to peak throughput)
  \label{tab:ours_vs_prior}%
    \begin{threeparttable}
    \begin{tabular}{|c|c|c|c|c|}
    \hline
    \multicolumn{1}{|p{4em}|}{\textbf{Precision (ip-op)}} & \multicolumn{1}{p{4em}@{}|}{\textbf{\gama TE\tnote{*}}} & \multicolumn{1}{p{5em}|}{\textbf{Best prior work} } &
    \multicolumn{1}{p{5em}|}{\textbf{Prior work TE\tnote{+}} } &
    \multicolumn{1}{p{5em}|}{\textbf{\% Improvement}} \bigstrut\\
    \hline
    \textbf{int8-int32} & 69\%  & MAXEVA\cite{taka:2023:fpt:maxeva} & 60\% & 9\% \bigstrut\\
    \hline
    \textbf{int8-int16} & 82\%  & AMA\cite{deng:2024:fpl:ama} & 73.3\% & 8.7\% \bigstrut\\
    \hline
    \textbf{int8-int8} & 85\% & AutoMM\cite{zhuang:2023:fpga:automm} & 31.3\%  & 53.6\% \bigstrut\\
    \hline
    \textbf{int8-int8} & 85\% & ARIES\cite{zhuang:2025:fpga:aries} & 45.9\% & 39\% \bigstrut\\
    \hline
    \textbf{bf16-bf16} & 86\% & - & -     & - \bigstrut\\
    \hline
    \end{tabular}%
    \begin{tablenotes}
    \item[*] AMD Versal VE2802 (VEK280 Board)
    \item[+] AMD Versal VC1902 (VCK190 Board)
    \end{tablenotes}
    \end{threeparttable}
\vspace{-6mm}
\end{table}%

\vspace{-2mm}
\subsection{Performance of a pack of AIEs}
% \vspace{-1mm}
Figure \ref{fig:aie_pack_4} shows the AIE connected as a pack. Each AIE in the pack uses the kernel size from the previous section, optimizing compute to communication ratio \(\gamma\), AIE memory utilization, and KCE.
We use a pack size of 4 as discussed in Section \ref{sec:gama_frame:pack}.
Table \ref{tab:pack_buff_plmt_res} shows the effect of different placement constraints on the pack. 
% impact of our custom buffer address placement on the pack of AIE.
Unconstrained buffers is considered as the baseline with the highest average KCE of 88\%. 
Buffers constrained to the pack using buffer location placement show 14\% loss in KCE.
Our custom buffer placement recovers 12\% of the loss, bringing the average KCE to 86\%.

Table \ref{tab:pack_buff_plmt_res} also shows the efficiency loss incurred as a result of the cascade interface. 
We use the KCC of unconstrained single AIE buffer from Table \ref{tab:sa_buff_plmt_res_mmul} as baseline to compare with the KCC of unconstrained buffer cycles of the pack. 
Since unconstrained buffer placement avoids any memory stalls,
the efficiency loss can be characterized as a loss due to cascade. 
% Since both cycle measurements are using an unconstrained buffer placement that avoids any memory stalls 
% Thus, the difference between those can be quantified as cascade cycles. 
On average, we see a 7\% reduction in KCE due to cascade stalls, compared to single AIE case, across all precisions.
% We see on average 8\% recovery in performance. 
% Our custom buffer placement which is confined to the pack helps us easily scale the design to the complete array.
% We can have different number of \aieml in the pack. Table \ref{tab:pack_size_variation} shows the trade-off of having a small pack versus a large pack. 
% Ideally, you can have any size of pack until you have the required PLIO resources to support the pack. 
% The main issue arises when you have to scale the pack to the complete AIE array for best performance. 

\vspace{-2mm}
\subsection{Scaling to the complete array}
%Versal VE2802 contains 304 AIE-ML tiles arranged in an 8 by 38 matrix.
With custom buffer address placement, resulting in buffers being placed within a pack, we are able to achieve easy and efficient scaling across the whole AIE array. 
The hyperparameter values of 8, 4 and 9 for Y, G and X, respectively, show the best scaling on VE2802.
% We use 4 AIEs in a single pack. 
% This pack is then replicated over the entire \aieml array as shown in Figure \ref{fig:aie_array_scaling}. 
Table \ref{tab:aie_array_final_perf} shows the final throughput achieved for the complete \aieml array for different precisions. 
Our design achieves 94.7\% (288/304) AIE array utilization, 94.7\% of memory bank utilization, 60\% PLIO in utilization and 85.7\% PLIO out utilization. 
We calculate execution time from timestamps generated by \texttt{aiesimulator} in the output file.
This throughput measurement technique was introduced by \maxeva \cite{taka:2023:fpt:maxeva} and is also used by AMA \cite{deng:2024:fpl:ama}, keeping the results consistent. 
The highest throughput of 165 TOPs is observed for int8-int8 due to low communication latency due to 8-bit precision.
Similarly, int8-int32 has the lowest throughput of 133 TOPs, the larger output precision occupies more memory and increases communication latency.  

% Table generated by Excel2LaTeX from sheet 'Results section '
% \begin{table*}[t]
%   \centering
%   \caption{Performance for our designs scaled to the whole AIE array}
%     \begin{tabular}{|l|c|c|c|r|r|r|r|l|}
%     \hline
%     \multicolumn{1}{|p{4em}|}{Precision} & \multicolumn{1}{p{1em}|}{Y} & \multicolumn{1}{p{1em}|}{G} & \multicolumn{1}{p{1em}|}{X} & \multicolumn{1}{p{2em}|}{AIEs} & \multicolumn{1}{p{3em}|}{Memory Banks} & \multicolumn{1}{p{2em}|}{PLIOs In } & \multicolumn{1}{p{2em}|}{PLIOs out} & \multicolumn{1}{p{3em}|}{Throughput (TOPs)}  \bigstrut\\
%     \hline
%     int8\_int32 & 8     & 4     & 9     & 288(94.7\%) & 2304 (94.7\%) & 68(60.7\%) & 72(85.7\%) & 133 (69\%)   \bigstrut\\
%     \hline
%     int8\_int16 & 8     & 4     & 9     & 288(94.7\%) & 2304 (94.7\%) & 68(60.7\%) & 72(85.7\%) & 159 (82\%)    \bigstrut\\
%     \hline
%     int8\_int8 & 8     & 4     & 9     & 288(94.7\%) & 2304 (94.7\%) & 68(60.7\%) & 72(85.7\%) & 165 (85\%)    \bigstrut\\
%     \hline
%     bf16\_bf16 & 8     & 4     & 9     & 288(94.7\%) & 2304 (94.7\%) & 68(60.7\%) & 72(85.7\%) & 83 (TBFLOPs) (86\%)    \bigstrut\\
%     \hline
%     \end{tabular}%
%   \label{tab:aie_array_final_perf}%
% \end{table*}%

\subsection{Comparing with other state-of-the-art designs}
\rev{Our work is the first work to map GEMM operations on full AIE-ML array devices.} 
Prior work \cite{zhuang:2023:fpga:charm,taka:2023:fpt:maxeva,deng:2024:fpl:ama,wang:2025:arxiv:rsn_xnn} use AIE. Hence, we only compare the relative achieved performance with prior work. 
Relative comparison is done using Throughput Efficiency (TE) i.e. ratio of achieved throughput to peak throughput.
% the total throughput the AIE array design can achieve compared to the peak performance of the particular chip. 
Moreover, different prior works use different precisions - some use int8-int8 \cite{zhuang:2023:fpga:charm,zhuang:2024:trets:charm_2,zhuang:2023:fpga:automm,zhuang:2025:fpga:aries}, int8-int16 \cite{deng:2024:fpl:ama}, int8-int32 \cite{taka:2023:fpt:maxeva}.
No other prior work use all the 4 precisions like GAMA. Hence, we can only compare a part of our results with each prior work. 
Table \ref{tab:ours_vs_prior} compares our approach with \cite{deng:2024:fpl:ama,zhuang:2023:fpga:charm,zhuang:2023:fpga:automm,zhuang:2025:fpga:aries} different works on Versal VC1902.

We compare int8 input and int32 output with MAXEVA \cite{taka:2023:fpt:maxeva}, since that is the only work that writes 32-bit output back to the PL. 
\gama achieves 133 TOPs (69\%) TE.
\maxeva achieves 77 TOPs (60\%) TE. 
\gama extracts 9\% more TE. MAXEVA's efficiency is capped at 80\% as only 80\% of AIEs perform matrix multiplication, with 20\% used for addition.
\begin{comment}
\red{
MAXEVA uses a separate add kernel for reduction and buffer data transfer method for communicating the partial sum, whereas our design uses a modified version of matrix multiplication kernel which supports reduction.
We use a cascade interface to communicate the partial sum from one \aieml to the next \aieml. 
Both methods have different tradeoffs. 
MAXEVA’s array efficiency is limited to 80\% since only 80\% of the AIEs in the design do matrix multiplication operation (20\% are used for addition).
The AIEs that perform the reduction help MAXEVA to have more room for buffer placement, reducing memory stalls. 
Our design, on the other hand, can use 94\% of the \aieml for the matrix multiplication operation. 
But since all the kernels do matrix multiplication, the design becomes tight, leading to high memory stalls.
Our custom buffer placement helps alleviate memory stalls as seen in Table \ref{tab:sa_buff_plmt_res_mmul} and improves the overall efficiency of the design.}
\end{comment}

AMA \cite{deng:2024:fpl:ama} uses a similar kernel design compared to ours. Their matrix multiplication kernels also perform the reduction. 
They communicate partial sums to the next engine using buffer connections. 
AMA only writes 16 bit outputs compared to MAXEVA's 32 bit output.
Comparing the TE of the chip, we are 8.7\% higher than AMA for the 16-bit output.  

Lastly, we compare \gama with CHARM \cite{zhuang:2023:fpga:automm} and ARIES \cite{zhuang:2025:fpga:aries} for int8-int8 precision. 
Our design outperforms CHARM and ARIES by 53.6\% and 39\%, respectively, in terms of TE. 
% While both designs have implemented the PL and demonstrated full hardware execution, our approach focuses on evaluating the performance of \aieml in simulation, providing valuable insights and paving the way for future hardware deployment.
RSN-XNN \cite{wang:2025:arxiv:rsn_xnn} only demonstrates performance for fp32. 
Since AIE-ML lacks native support for fp32, a direct comparison between \gama and RSN-XNN cannot be made.
% MENTION RXN paper.. Say it's only FP32
 
% The default compiler flags, and buffer optimization do not do a good job when it comes to buffer placement for a tight design. To avoid the memory stall, we place the buffers manually in the design leading to reduced memory stalls.
\section{Conclusion}
% First work on AIE2. 
In this work, we introduce the \gama framework that accelerates GEMM on AMD Versal AIE-ML architecture. 
Our framework maximizes AIE memory utilization, provides a custom buffer placement algorithm to minimize memory stalls, and a staggered kernel placement to facilitate scaling over the AIE array. 
\gama framework provides the highest utilization of peak throughput for Versal compared to all other state-of-the-art frameworks. 
\rev{Though our improvements focus on AIE-ML, optimizing memory usage and prioritizing scalability can also benefit other accelerator architectures.}
% Maximizing AIE mem util, cascade, packing, scaling to the array, buffer placement gives higher performance. 

%\textcolor{red}{Future directions.
% We leave incorporating memory tiles and optimizing PL buffers for future work. 
% Both these memories can act as local storage for GEMM applications. 
% Maximizing utilization of these memories is critical to improving the overall efficiency due to Versal's limited peak off-chip DRAM bandwidth.
%Memory tiles can be used as local storage for GEMM  tasks. Making the best use of these tiles is very important because Versal devices have limited off-chip DRAM bandwidth.
%Improving their usage can significantly boost overall performance. However, optimizing these memories will be addressed in future work.

% to maximize performance memory utilization

Future work will incorporate the memory tiles present in the AIE-ML architecture. 
Maximizing utilization of these memories is critical to improving the overall performance due to Versal's limited peak off-chip DRAM bandwidth.
% For PL memory optimization a systematic methodology can be exploited similar to cite{FCCM paper goes here, https://ieeexplore.ieee.org/abstract/document/10653656} for AIE1. However, MemTiles introduce a new challenge, which we leave as future work. 

% \section*{Acknowledgment}

% The preferred spelling of the word ``acknowledgment'' in America is without 
% an ``e'' after the ``g''. Avoid the stilted expression ``one of us (R. B. 
% G.) thanks $\ldots$''. Instead, try ``R. B. G. thanks$\ldots$''. Put sponsor 
% acknowledgments in the unnumbered footnote on the first page.
\bibliographystyle{IEEEtranS}
\bibliography{refs}

\vspace{12pt}

\end{document}